%% file: main.tex
\documentclass[
reprint,
superscriptaddress,
frontmatterverbose, 
showpacs,
preprintnumbers,
nofootinbib,
amsmath,
amssymb,
floatfix,
twocolumn
]{revtex4-2}

\usepackage{soul}
\usepackage[inline]{enumitem}
\usepackage[utf8]{inputenc}
\usepackage[normalem]{ulem}
\usepackage{graphicx}% Include figure files
\usepackage{dcolumn}% Align table columns on decimal point
\usepackage{bm}% bold math
\usepackage{color}
\usepackage[dvipsnames]{xcolor}
\usepackage[
    colorlinks = true,
    linkcolor = BlueViolet,
    anchorcolor = purple,
    citecolor = purple,
    filecolor = purple,
    urlcolor = BlueViolet]{hyperref}
    
\usepackage{url}
\usepackage{xspace}
\usepackage{slashed}
\usepackage{multirow,bigstrut}
\usepackage{mathrsfs} % pretty maths
\usepackage{relsize,amsmath}
\usepackage[geometry]{ifsym}
\usepackage{amssymb}% http://ctan.org/pkg/amssymb
\usepackage{pifont}% http://ctan.org/pkg/pifont
\usepackage{physics}

\usepackage{cleveref}
\usepackage{rotating}
\usepackage{ragged2e}
\usepackage{lipsum}

\usepackage{fontawesome} % for code link icons
\definecolor{blue-violet}{rgb}{0.33, 0.17, 0.89}

\newcommand{\titleprl}[1]{\vspace{0.3cm} \noindent \textit{\textbf{#1}}:}
\input{custom_commands.tex}
\usepackage{siunitx}
\input{units}

\begin{document}

\title{\Large Dark Matter phase space distribution under a fully relativistic treatment \\
         \large \it Eddington inversion according to general relativity }

% AUTHORS
%%%%%%%%%%%%%%%%%%%%%%%%%%%%%%%%%%%%%%%%%%%%%%%
%%%%%%%%%%%%%%%%%%%%%%%%%%%%%%%%%%%%%%%%%%%%%%%
\author{Jaime Hoefken Zink}
\email{jaime.hoefkenzink@ncbj.gov.pl}
\affiliation{National Centre for Nuclear Research, Pasteura 7, Warsaw, PL-02-093, Poland} 
%%%%%%%%%%%%%%%%%%%%%%%%%%%%%%%%%%%%%%%%%%%%%%%
\author{Pellegrino Piantadosi}
\email{pellegrino.piantadosi@nyu.edu}
\affiliation{New York University Abu Dhabi, PO Box 129188, Saadiyat Island, Abu Dhabi, UAE} 

%%%%%%%%%%%%%%%%%%%%%%%%%%%%%%%%%%%%%%%%%%%%%%%

\date{\today}

\begin{abstract}
We develop a fully relativistic extension of Eddington inversion to reconstruct the phase-space distribution function of dark matter from its density profile. Treating dark matter as an effective perfect fluid in a static, spherically symmetric spacetime, we use the Tolman-Oppenheimer-Volkoff equations to determine the gravitational background and derive an Abel-type inversion that yields the distribution function in terms of the conserved particle energy measured at infinity. We show that the resulting expression recovers the classical Eddington inversion in the weak-field, nonrelativistic limit. We also investigate the finite-boundary contribution, which can introduce an unphysical divergence, and present a regularization procedure that removes it. We apply our method to a Milky Way model consisting of a generalized Navarro-Frenk-White halo and a Gondolo-Silk dark-matter spike around Sagittarius A$^*$, neglecting baryonic contributions. The relativistic and Newtonian velocity distributions agree at radii far from the Galactic center but begin to differ at distances of order $10^{-5}\,\mathrm{pc}$. Closer to the black hole, the relativistic peak speed can reach approximately twice the Newtonian prediction. Finally, we derive a general expression for dark-matter capture by neutron stars that accommodates arbitrary ambient velocity distributions. Near Sagittarius A$^*$, the relativistic capture factor is nearly an order of magnitude smaller than the Maxwell-Boltzmann prediction, while the Newtonian Eddington result can overestimate it by a factor of approximately two. These results demonstrate the importance of relativistic phase-space reconstruction for dark-matter phenomenology in strong gravitational fields.
\end{abstract}

\maketitle

%%%%%%%%%%%%%%%%%%%%%%%%%%%%%%%
%        INTRODUCTION
\section{Introduction}

The Standard Model of particles physics (SM) is one of the most accurate scientific theories in predicting the observable reality that surround us. However, it cannot account for what we believe is one of the most important components of matter in the universe: dark matter (DM). Although we still do not know whether this component is made of particles, this is one of the strongest hypothesis. Several experiments have been built in order to detect DM particles (see for instance \cite{Cirelli:2024ssz}). Most of these observations highly depend on how DM behaves around us and, therefore, knowledge of its distribution is of extreme importance. Accurate knowledge of the local and global DM velocity distribution is essential for predicting recoil rates in direct detection experiments~\cite{Feldstein:2014gza}, calculating annual modulation signatures~\cite{Freese:2012xd}, evaluating DM capture rates in stellar bodies~\cite{Bramante:2017xlb, Busoni:2017mhe, Bell:2020jou, Leane:2022hkk}, and determining flux predictions for indirect searches sensitive to velocity-dependent annihilation processes (e.g., $p$-wave~\cite{Johnson:2019hsm, Piccirillo:2022qet}, $d$-wave~\cite{Piccirillo:2022qet}, or Sommerfeld-enhanced scattering~\cite{Facchinetti:2022jtd}).

A precise phase-space distribution function (PSDF) and corresponding velocity distribution function (VDF) of DM particles are critical ingredients across particle astrophysics and cosmology. Historically, DM phenomenological studies often adopted the Standard Halo Model (SHM)—an isotropic Maxwell-Boltzmann speed distribution truncated at an \textit{ad hoc} escape velocity~\cite{Mayet:2016zxu}. However, such an Ansatz lacks dynamical self-consistency in realistic, non-isothermal galactic potentials~\cite{Lacroix:2018qqh}, making the rigorous reconstruction of the DM velocity distribution directly from physical mass and potential models a central endeavor in particle astrophysics.

This is precisely what Eddington found possible~\cite{eddington1916distribution}: a reconstruction of the PSDF from the distribution of DM (and matter) across the galaxy. This method has widely been used in the literature~\cite{Ullio:2000bf, Vergados:2002hc, Catena:2011kv, Pato:2012fw, Bhattacharjee:2012xm, Bozorgnia:2013pua, Fornasa:2013iaa, Ferrer:2013cla, Hunter:2013vua, Lavalle:2014rsa, Boddy:2017vpe, Petac:2018gue, Lacroix:2018qqh, Nunez-Castineyra:2019odi, Lacroix:2020lhn, Lopes:2020dau, Christy:2023mgs, Almeida:2024qmi, Herrera:2026hal, Das:2026jsq}.
The method was built from a classical weak gravitational field approach. This may be completely precise for local DM and even for regions closer to the galactic center. However, as we get closer to Sgr A$^*$, the Newtonian description of gravity ceases to be valid. In the two last decades there have been an gradually increasing interest in testing DM at the galactic center, since there may be a much dense region, usually identified as a \textit{dark matter spike}~\cite{Lacroix:2016qag, John:2023knt, Hu:2023ubk, Aschersleben:2024xsb, Gustafson:2025ypo, Acevedo:2025rqu, Meighen-Berger:2026idy}. Although it has been difficult to get this spike from \textit{ab initio} simulations, conserved quantities in the emergence of our super-massive black hole, Sagittarius A$^*$, and numerous studies on the orbits of stars very close to the center of the Milky Way suggest the DM spike may be a real phenomenon~\cite{Becerra-Vergara:2020xoj, Becerra-Vergara:2021gmx, Nampalliwar:2021tyz, Chan:2022nra, Shen:2023kkm, DellaMonica:2023dcw, GRAVITY:2024tth, Crespi:2025ygl, Li:2025qcv, Yu:2026ebx}.

Therefore, a new way of computing PSDF should be found if we ever want to estimate with more precision interactions for which DM in the very galactic center is involved. This is precisely the aim of the present study: to develop a self-consistent method fully under the Einsteinian description of gravity, such that the PSDF of DM can be found directly from its density profile. We will divide this task in three sections. In \Cref{sec:EddInvS}, we will present the classical and relativistic Eddington inversion procedures. We will also discuss the presence of a boundary term that may have an unphysical contribution and show how to compute the velocity distributions in under Eddington inversion and by using a Maxwell - Boltzmann approximation. In \Cref{sec:DMMW}, we will present our choice of DM density profile in the Milky Way, which is the Navarro - Frenk- White with a Gondolo - Silk spike and will discuss the computations introduced in the previous section for that profile. We will compare the results and extract the region where the relativistic approach is relevant. Finally, we will summarize our results in \Cref{sec:Conc}.

%%%%%%%%%%%%%%%%%%%%%%%%%%%%%%%
%     EDDINGTON INVERSION
\section{Eddington inversion: a fully relativistic approach}
\label{sec:EddInvS}

The main aim of Eddington inversion is to be able to reconstruct the phase-space distribution of DM from its density profile. We will first show the original approach and then present a new one where the full general relativity approach is taken into account, for an isotropic distribution of DM.

\subsection{Classical Eddington inversion}
\label{sec:EddInv}

The foundational mathematical framework for reconstructing phase-space densities from density profiles was established by Eddington \cite{eddington1916distribution}. For a spherically symmetric, ergodic system in dynamical equilibrium, Eddington showed that a given mass density profile $\rho(r)$ embedded in a total gravitational potential $\Psi(r)$ uniquely determines an isotropic phase-space distribution function $f(E)$, where $E = \Psi(r) - v^2/2$ is the relative binding energy per unit mass. By integrating $f(E)$ over velocity space, one can directly derive a self-consistent velocity distribution function $f(v)$ rooted in the underlying gravitational potential of the galaxy.

Over the past two decades, Eddington's inversion method has become a cornerstone technique for reconstructing self-consistent dark matter velocity distributions and replacing Maxwellian approximations for direct detection calculations and in astrophysical scenarios~\cite{Ullio:2000bf, Vergados:2002hc, Catena:2011kv, Pato:2012fw, Das:2026jsq}. Establishing self-consistent velocity profiles exhibit significant departures from a simple Maxwellian shape, so that reconstructed non-Maxwellian VDFs alter direct detection exclusion bounds—particularly for light WIMPs and inelastic dark matter models. The results can deviate from standard Maxwell-Boltzmann models by orders of magnitude near energy thresholds~\cite{Bhattacharjee:2012xm}.

The method consists in reconstructing the energy distribution function for DM in terms of the negative relative energy per unit mass, $\epsilon \equiv \Psi(r) - v^2 / 2$, where $v$ is the velocity of the DM particle and $\Psi(r) \equiv \Phi_0 - \Phi (r)$ is the positive relative gravitational potential, found from the gravitational potential, $\Phi (r)$. $\Psi(r)$ can be directly calculated as,
\begin{equation}
    \Psi (r) = \int_r^{R_\mathrm{max}} dr' \frac{G m(r')}{r^{\prime 2}}\,,
\end{equation}
where $G$ is Newton's gravitational constant and $m(r)$ the total mass in the galaxy from $0$ to $r$. With this definition, $\Phi_0 \equiv \Phi(R_\mathrm{max})$. It is important to stress that the potential is computed from all the mass in the galaxy. However, in most of the regions we can approximate it to the DM content and the super-massive BH and, in the presence of a DM spike in the center of the galaxy, we can do this everywhere.

The DM density is equal to $\rho_\mathrm{DM} = \int d^3 \vec{v} \, f(\epsilon, L)$, where $f(\epsilon, L)$ is the desired PSDF. In our case, since we are considering spherically symmetric and isotropic DM, $f(\epsilon) \equiv f(\epsilon, L)$. We can express this equality in terms of the positive energy per mass $\epsilon$,
\begin{equation}
\rho_\mathrm{DM} (r) = 4 \pi \sqrt{2} \int_0^{\Psi (r)} f(\epsilon) \sqrt{\Psi (r) - \epsilon} \, d\epsilon\,,
\end{equation}
where the lower limit comes from the fact that we are just considering self-gravitating matter, so that $\epsilon > 0$. We can differentiate the density with respect to $\Psi$ to obtain:
\begin{equation}
\frac{d\rho_\mathrm{DM}}{d\Psi} = \sqrt{8} \pi \int_0^{\Psi} \frac{f(\epsilon)}{\sqrt{\Psi - \epsilon}} \, d\epsilon\,.
\end{equation}
This takes the precise form of an Abel equation, whose solution is,
\begin{equation}
f(\epsilon) = \frac{1}{\sqrt{8} \pi^2} \frac{d}{d\epsilon} \int_0^{\epsilon} \frac{d\Psi}{\sqrt{\epsilon - \Psi}} \frac{d\rho_\mathrm{DM}}{d\Psi}\,.
\end{equation}

By integrating by parts, we can get rid of the derivative,
\begin{equation}
\label{eq:EddInvNR}
f(\epsilon) = \frac{1}{\sqrt{8} \pi^2} \left[\frac{1}{\sqrt{\epsilon}} \left( \frac{d\rho_\mathrm{DM}}{d\Psi} \right)_{\Psi=0} +  \int_0^{\epsilon} \frac{d\Psi}{\sqrt{\epsilon - \Psi}} \frac{d^2\rho_\mathrm{DM}}{d\Psi^2} \right]\,.
\end{equation}

As $R_\mathrm{max} \to \infty$, then the boundary term goes to $0$. When we integrate from to a finite value of $R_\mathrm{max}$, then the boundary term may contribute with a diverging value that alters the PSDF. More precise studies have also analyzed how to redefine the parameters so that this situation is circumvented~\cite{Lavalle:2014rsa}, though in the literature this term is usually dropped~\cite{Yang:2024jtp}.

\subsection{Fully relativistic Eddington inversion}
\label{sec:GREddInv}

In order to find a suitable expression for the PSDF suitable for general relativity, we will use the time component of a metric of the form:
\begin{equation}
ds^2 = - g_{tt} dt^2 + g_{rr} dr^2 + r^2 d\Omega^2\,.
\end{equation}

We need to relate the density to $g_{tt}$. We will assume DM behaves like a perfect fluid and will just regard the effect of it and of Sgr A$^*$.

By considering another form of the TOV equations, it is possible to reconstruct the metric element throughout the galaxy, just as an effect of DM. We will assume we know the density, $\rho (r)$. The mass is easily reconstructed as,
\begin{equation}
\begin{split}
\frac{dM(r)}{dr} &= 4 \pi r^2 \rho(r)\,.
\end{split}
\end{equation}
It is imporant to take into account that when this equation is solved, we need to add $M_\mathrm{BH}$, the mass of Sgr A$^*$, to the total mass, so that we consider the BH mass as part of the total enclosed mass.
With this value, we can also obtain the pressure $P$ and the metric element $g_{tt}$, under the assumption of static spherical symmetry. This is encoded in another form of the TOV equations,  
\begin{equation}
\begin{split}
\frac{dP(r)}{dr} &= -\frac{\rho (r) + P(r)}{r^2 \left( 1 - \frac{2 G M(r)}{r} \right)} \left( 4\pi r^3 P(r) + G M(r) \right)\,,\\
\frac{dg_{tt}(r)}{dr} &= \frac{2 g_{tt} (r)}{r^2} \left( 4\pi r^3 P(r) + G M(r) \right) \left[ 1 - \frac{2 G M(r)}{r} \right]^{-1}\,.
\end{split}
\end{equation}

This system can be solved under the following boundary conditions,
\begin{equation}
\begin{split}
g_{tt} \left(r_\mathrm{max}\right) &= 1 - \frac{2 G M \left( r_\mathrm{max} \right)}{r_\mathrm{max}}\,,\\
P \left(r_\mathrm{max}\right) &= 0\,,
\end{split}
\end{equation}
where $r_\mathrm{max}$ is the minimum radius at which the density can be assumed to vanish. As standard approaches when Eddington inversion is performed, we will neglect the effect of the boundary term.

We can take the equation for the density in terms of the distribution function, and recast it in terms of $\mathcal{E} \equiv \sqrt{g_{tt}} \, E_\mathrm{DM}^\mathrm{loc} / m_\mathrm{DM}$, the coordinate conserved energy, and $g_{tt}$. Here, $E_\mathrm{DM}^\mathrm{loc}$ is the local energy of a DM particle of mass $m_\mathrm{DM}$. Then, we can define the following quantity $\mathcal{F} \equiv \rho g_{tt}^2$, such that,
\begin{equation}
\begin{split}
\mathcal{F} &= 4 \pi m_\mathrm{DM}^4 \int_{\sqrt{g_{tt}}}^1 d\mathcal{E} \, f(\mathcal{E}) \, \mathcal{E}^2 \sqrt{\mathcal{E}^2 - g_{tt}}\,,
\end{split}
\end{equation}
where the integration is up to $1$ just to consider bounded DM particles.
We can take the derivative of $\mathcal{F}$ as a function of $g_{tt}$ to get,
\begin{equation}
\label{eq:dF_dgtt}
\begin{split}
\frac{d\mathcal{F}}{d g_{tt}} &= - 2 \pi m_\mathrm{DM}^4 \int_{\sqrt{g_{tt}}}^1 d\mathcal{E} \, \frac{f(\mathcal{E}) \, \mathcal{E}^2}{\sqrt{\mathcal{E}^2 - g_{tt}}}\,.
\end{split}
\end{equation}

Now we can perform the following change of variables:
\begin{equation}
\begin{split}
\mathcal{H} &\equiv 1 - \mathcal{E}^2\,,\\
h\left(\mathcal{H}\right) &\equiv f\left(\sqrt{1 - \mathcal{H}}\right) \sqrt{1 - \mathcal{H}}\,,\\
\chi &\equiv 1 - g_{tt}\,.
\end{split}
\end{equation}

Then, \Cref{eq:dF_dgtt} takes the following form,
\begin{equation}
\label{eq:dJ_dchi}
\begin{split}
\frac{1}{\pi m_\mathrm{DM}^4}\frac{d\mathcal{F}}{d \chi} &= \int_{0}^\chi d\mathcal{H} \, \frac{h\left(\mathcal{H}\right)}{\sqrt{\chi - \mathcal{H}}}\,.
\end{split}
\end{equation}

\Cref{eq:dJ_dchi} can be solved for $h\left(\mathcal{H}\right)$ by means of Abel's integral formula, so that we find,
\begin{equation}
\label{eq:h_H}
\begin{split}
h\left(\mathcal{H}\right) = \frac{1}{\pi^2 m_\mathrm{DM}^4} \frac{d}{d \mathcal{H}} \int_0^\mathcal{H} \frac{d\mathcal{F}}{d\chi} \frac{d\chi}{\sqrt{\mathcal{H} - \chi}}\,.
\end{split}
\end{equation}

This solution can be written in terms of our original variables,
\begin{equation}
\label{eq:f_solution_0}
\begin{split}
f\left(\mathcal{E}\right) = \frac{1}{2 \pi^2 m_\mathrm{DM}^4 \mathcal{E}^2} \frac{d}{d \mathcal{E}} \int_{\mathcal{E}^2}^1 \left( g_{tt}^2 \frac{d\rho}{d g_{tt}} + 2 \rho g_{tt} \right) \frac{dg_{tt}}{\sqrt{g_{tt} - \mathcal{E}^2}}\,.
\end{split}
\end{equation}

By integrating by parts , we find the final form for the distribution function:
\begin{equation}
\label{eq:f_solution}
\begin{split}
f\left(\mathcal{E}\right) = &\frac{1}{\pi^2 m_\mathrm{DM}^4 \mathcal{E}} \Bigg[ \left( \frac{d\left( g_{tt}^2 \rho \right)}{d g_{tt}} \right)_{r \to r_\mathrm{max}} \times \frac{1}{\sqrt{1 - \mathcal{E}^2}} \\&+\int_{\mathcal{E}^2}^1 \left( 2 \rho + 4 \, g_{tt} \frac{d\rho}{d g_{tt}} + g_{tt}^2 \frac{d^2\rho}{d g_{tt}^2} \right) \frac{dg_{tt}}{\sqrt{g_{tt} - \mathcal{E}^2}} \Bigg]\,,
\end{split}
\end{equation}
where the first term in square brackets is a boundary term. When $r_\mathrm{max}$ is sufficiently large, then we expect the expression to vanish. However, for computations where we do not go to infinity and when we consider a finite escape velocity on the boundaries, this term can generate a non-physical divergence, such as it happens to the non-relativistic Eddington inversion.   

By solving the TOV equations, we can find both $\rho$ and $g_{tt}$ as a function of $r$, from which we can construct the function $\rho \left( g_{tt} \right)$, which can be used to solve \Cref{eq:f_solution}.

\subsection{Newtonian Limit}
An important consistency check of the relativistic inversion derived above is that it reproduces the classical Eddington inversion in the weak-field and non-relativistic limits, $\Psi \ll1$, $v^2 \ll 1$, in natural units ($c = 1$). For a static weak gravitational field, the time component of the metric can be written as (assuming $r_{\text{max}}\gg r_s$),  
\begin{equation}
    g_{tt}(r) \approx 1 - 2\Psi (r).
\end{equation}
Similarly, the relativistic conserved energy per unit of rest mass, $E=\sqrt{g_{tt}}\left(1-v^2\right)^{-1/2}$, becomes
\begin{align}  
    E&\approx\left(1-\Psi\right)\left(1+\frac{v^2}{2}\right) \\
    &\approx 1-\left(\Psi-\frac{v^2}{2}\right) \\
    &\approx 1-\epsilon \ .
\end{align}
In the weak-field limit, $E^2\approx1-2\epsilon$ and $g_{tt}\approx1-2\Psi$, consequently,
\begin{equation}\label{eq:newton_limit1}
    g_{tt}-E^2 \approx 2\left(\epsilon-\Psi\right)\,,
\end{equation}
while $dg_{tt}\approx-2\, d\Psi$. Moreover,
\begin{equation}\label{eq:newton_limit2}
    \mathcal{F}\approx\rho \ \Rightarrow \frac{d\mathcal{F}}{dg_{tt}}\approx-\frac{1}{2}\frac{d\rho}{d\Psi}\,.
\end{equation}
Substituting (\ref{eq:newton_limit1}),(\ref{eq:newton_limit2}) in (\ref{eq:f_solution_0}), one obtains, at leading order,
\begin{equation}
    f_{GR}(1-\epsilon) \rightarrow \frac{1}{2\sqrt{2}\pi^2m^4_{DM}}\frac{d}{d\epsilon}\int_0^{\epsilon}\frac{d\rho/d\Psi}{\sqrt{\epsilon-\Psi}}d\Psi \ .
\end{equation}
which, according to the normalization adopted in \ref{sec:GREddInv}, matches exactly with the distribution presented in Section \ref{sec:EddInv}.

\subsection{Boundary term}
\label{sec:regularization}

There is a divergence arising from the factor $\frac{1}{\sqrt{1 - \mathcal{E}^2}}$ in the boundary term. This could be circumvented by sending the boundary to infinity, however the boundary should not be zero due to neighboring galaxies and their gravitational effects. The first common approach is just to neglect this term. We will take this approach to compare the relativistic to the non-relativistic results. However, we can also regularize the parameters in such a way that we obtain a consistent Eddington inversion, while keeping the boundary term. Here, we will take a similar approach as with the classical inversion, by regularizing the density profile itself, such that the boundary term vanished identically for $r = r_\mathrm{max}$.

Following the strategy by \cite{Lacroix:2018qqh}, we also propose to modify the density profile:
\begin{equation}
\tilde{\rho} = \rho - \mathfrak{f}(g_{tt})\,,
\end{equation}
where $\tilde{\rho}$ is a new density profile and $\mathfrak{f}(g_{tt})$ is a function to be found. We want the new density profile to make the boundary term to vanish at $r = r_\mathrm{max}$. Therefore, the following expression
\begin{equation}
\begin{split}
\frac{d\left( g_{tt}^2 \tilde{\rho} \right)}{d g_{tt}} = g_{tt}^2 \frac{d \rho}{d g_{tt}} + 2 g_{tt} \rho - g_{tt}^2 \frac{d\mathfrak{f}}{d g_{tt}} - 2 g_{tt} \mathfrak{f}\,,
\end{split}
\end{equation}
has to vanish at the boundary. Since we are just interested in its value at the boundary, we can assume some of the parameters of this equation are in its general expression just evaluated at $r = r_\mathrm{max}$. So, we need to solve the following differential equation,
\begin{equation}
\begin{split}
\dot{\mathfrak{f}} + \frac{2}{{g_{tt}}_0} \mathfrak{f} = \dot{\rho}_0 + \frac{2}{{g_{tt}}_0}\rho_0 \,,
\end{split}
\end{equation}
where the subscript $0$ means that the quantity is evaluated at the boundary and the dotted quantities are derivatives with respect to $g_{tt}$. A general solution of the previous equation is,
\begin{equation}
\begin{split}
\mathfrak{f}(g_{tt}) = A \, \exp \left(- \frac{2}{{g_{tt}}_0} g_{tt} \right) + \frac{{g_{tt}}_0}{2} \dot{\rho}_0 + \rho_0 \,,
\end{split}
\end{equation}
where $A$ is a general constant.

Since the aim of the method is to deviate as least as possible, we can find the minimum of $\mathfrak{f}^2(g_{tt})$ with respect to $A$, in order to minimize the density deviation on the boundary. This gives us,
\begin{equation}
\begin{split}
A = - \frac{1}{2} e^2 \left( {g_{tt}}_0 \dot{\rho}_0 + 2 \rho_0 \right) \,.
\end{split}
\end{equation}

We will use this fixed value to regularize the density profile, so that its final form is,
\begin{equation}
\tilde{\rho} = \rho + \frac{1}{2} e^2 \left( {g_{tt}}_0 \dot{\rho}_0 + 2 \rho_0 \right)  \exp \left(- \frac{2}{{g_{tt}}_0} g_{tt} \right)  - \frac{{g_{tt}}_0}{2} \dot{\rho}_0 - \rho_0\,.
\end{equation}

It can easily be checked that $\left(d\left( g_{tt}^2 \tilde{\rho} \right) /d g_{tt} \right)_0 = 0$. Now, we can perform the computations with this new density profile and find a regularized solution for the PSDF,
\begin{equation}
\label{eq:f_solution_reg1}
\begin{split}
\tilde{f}\left(\mathcal{E}\right) = &\frac{1}{\pi^2 m_\mathrm{DM}^4 \mathcal{E}} \int_{\mathcal{E}^2}^1 \left( 2 \tilde{\rho} + 4g_{tt} \frac{d\tilde{\rho}}{d g_{tt}} + g_{tt}^2 \frac{d^2\tilde{\rho}}{d g_{tt}^2} \right) \frac{dg_{tt}}{\sqrt{g_{tt} - \mathcal{E}^2}}\,.
\end{split}
\end{equation}

\subsection{Velocity distribution}

With Eddington inversion method, we can find a distribution in terms of a positive classical energy per mass ($\epsilon$) or the general relativistic invariant energy per mass ($\mathcal{E}$). However, we are directly interested in the VDF. Since the density is a key ingredient in the computations, we are just interested in finding the form of the distribution, so that we will normalize it to $1$. Therefore, we will present the transformations regardless of constants that may appear, since they will be washed out by the rescaling.

For the classical case, since $f_{\vec{v}} (\vec{v}, r) \equiv f(\epsilon, L) / \rho(r)$~\cite{Lacroix:2018qqh}, where $f_{\vec{v}} (\vec{v}, r)$ is the distribution of the 3D-velocity, then $f_v (v, r) \propto v^2 f(\epsilon \equiv \Psi(r) - v^2/2)$, where $f_v (v, r)$ is the VDF at $r$ and we have integrated $L$ out due to the isotropic distribution considered.

The relativistic case is more subtle. Here, we have the following relation to local density, where we will drop any explicit dependence on $r$,
\begin{equation}
\begin{split}
\rho &\propto \int d\mathcal{E} \, f(\mathcal{E}) \, \mathcal{E}^2 \sqrt{\mathcal{E}^2 - g_{tt}}\,,\\
&\propto \int d^3 \vec{v} \, f_{\vec{v}} (\vec{v}) \equiv \int dv \, f_v (v)\,.
\end{split}
\end{equation}

Since $\mathcal{E} = \sqrt{g_{tt}} \gamma \equiv \sqrt{g_{tt}} / \sqrt{1 - v^2}$, then $f(\mathcal{E}) = f\left(\sqrt{1 - g_{tt} / \mathcal{E}^2}\right)$, $d\mathcal{E} = \sqrt{g_{tt}} \gamma^3 \, v \, dv$, and $\sqrt{\mathcal{E}^2 - g_{tt}} = \sqrt{g_{tt}} \gamma \, v$. Therefore,
\begin{equation}
\begin{split}
\rho &\propto \int dv \, \gamma^6 \, v^2 \, f\left(\sqrt{1 - g_{tt} / \mathcal{E}^2}\right)\,,\\
f_v (v) &\propto \gamma^6 \, v^2 \, f\left(\sqrt{1 - g_{tt} / \mathcal{E}^2}\right)\,.
\end{split}
\end{equation}

In this way, there is an extra factor of $\gamma^6$ that is not present in the classical framework.

\subsection{Maxwell - Boltzmann distribution}
\label{sec:MB}

In the standard approach to velocity distributions, DM has been treated as an ideal gas, exhibiting a Maxwell - Boltzmann velocity distribution~\cite{Busoni:2017mhe, Bell:2020jou}. The form this VDF takes is as follows,
\begin{equation}
f_\mathrm{MB} (v) = 4 \, \pi \left( \frac{3}{2 \pi v_d^2} \right)^{3/2} v^2 \exp \left( - \frac{3 v^2}{2 v_d^2} \right)\,,
\end{equation}
where $v_d$ is the DM 3D-velocity dispersion. The distribution peaks at $\sqrt{2/3} \, v_d$ and has a mean value of $\langle v \rangle_\mathrm{MB} = \sqrt{8/3\pi} \, v_d \sim 0.92 \, v_d$. In this 1D-distribution, the dispersion is $\sqrt{v_d^2 - \langle v \rangle_\mathrm{MB}^2} \sim 0.39 \, v_d$.

To find $v_d$ for different distances, we interpret it as the rotation velocity, $v_\mathrm{rot}$, as can be found in \cite{Shen:2023kkm},
\begin{equation}
v_\mathrm{rot} (r) = \sqrt{\frac{G \, M(r)}{r}}\,,
\end{equation}
where $G$ is Newton's gravitational constant and $M(r)$ is the total mass of the galaxy from the center to $r$. It can be found that this yields $v_\mathrm{rot} (r = 8 \, \mathrm{kpc}) \approx 270$ km/s, which is precisely the value used for $v_d$ in standard computations for local DM~\cite{Bell:2020jou}.

%%%%%%%%%%%%%%%%%%%%%%%%%%%%%%%
%        
\section{Dark matter distribution in the Milky Way}
\label{sec:DMMW}

In order to apply the previous method to the reconstruction of the PSDF of DM, we will consider a halo profile together with a spike close to Sgr A$^*$. These will be modeled with the general Navarro-Frenk-White and the Gondolo - Silk profiles respectively. In order to have a full proper computation, we would need to account also for the baryonic mass. However, this is out of the scope of the present work. Since we are interested in analyzing roughly the behavior of the relativistic Eddington inversion with respect to the classical one anywhere in the galaxy and, more accurately, the relativistic effects close to the super-massive BH, we do not need to fully describe baryonic matter, since close to Sgr A$^*$ just the BH and the DM spike are the main sources of gravitational effects.

\begin{figure*}
    \centering
    \includegraphics[width=0.49\linewidth]{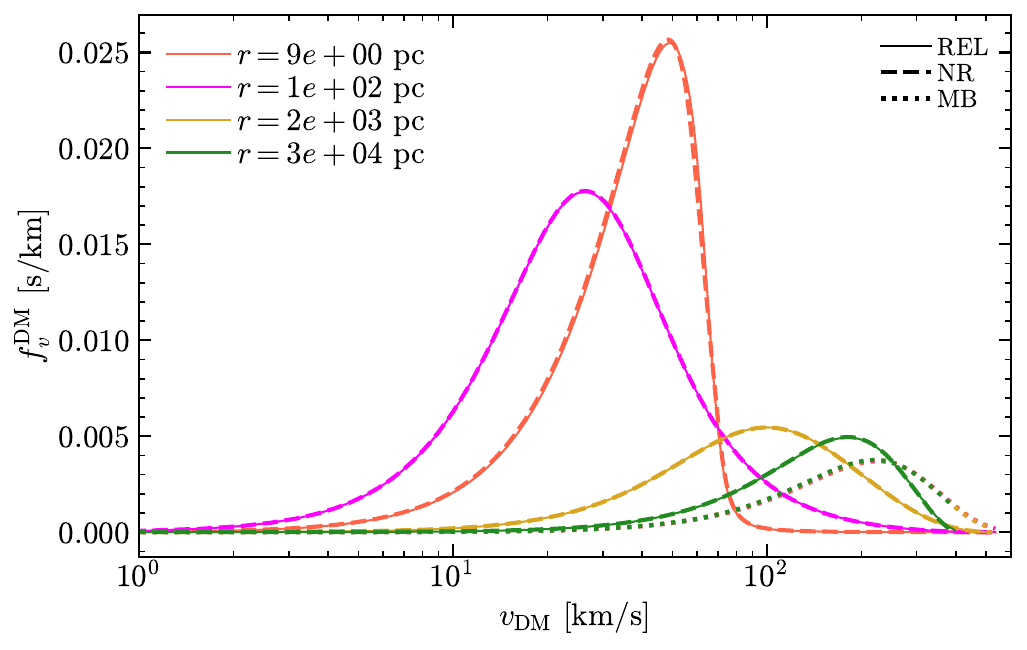}
    \includegraphics[width=0.49\linewidth]{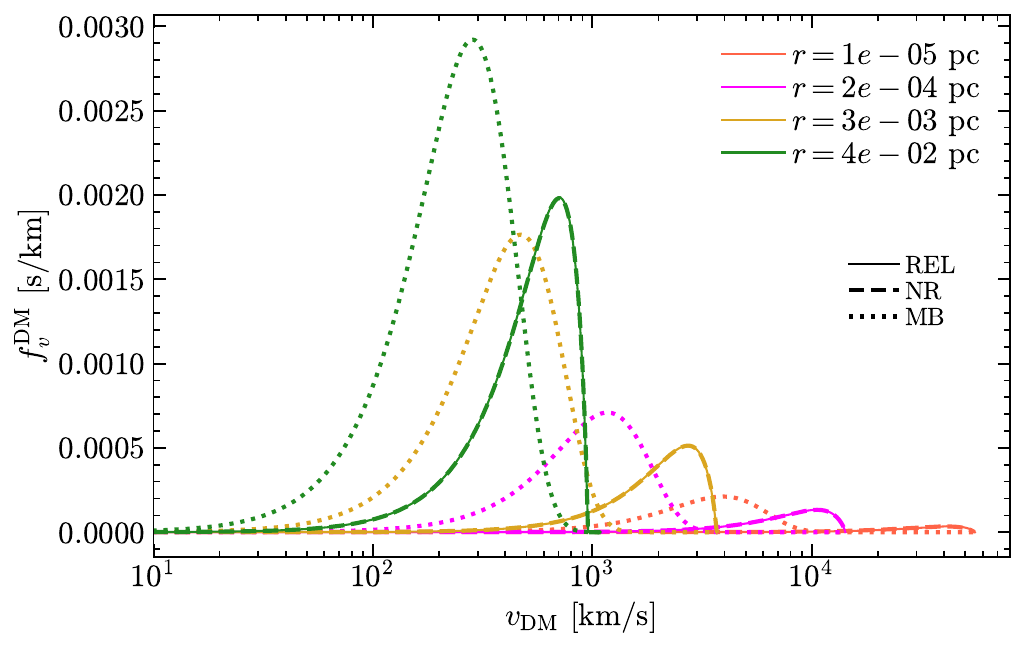}
    \includegraphics[width=0.49\linewidth]{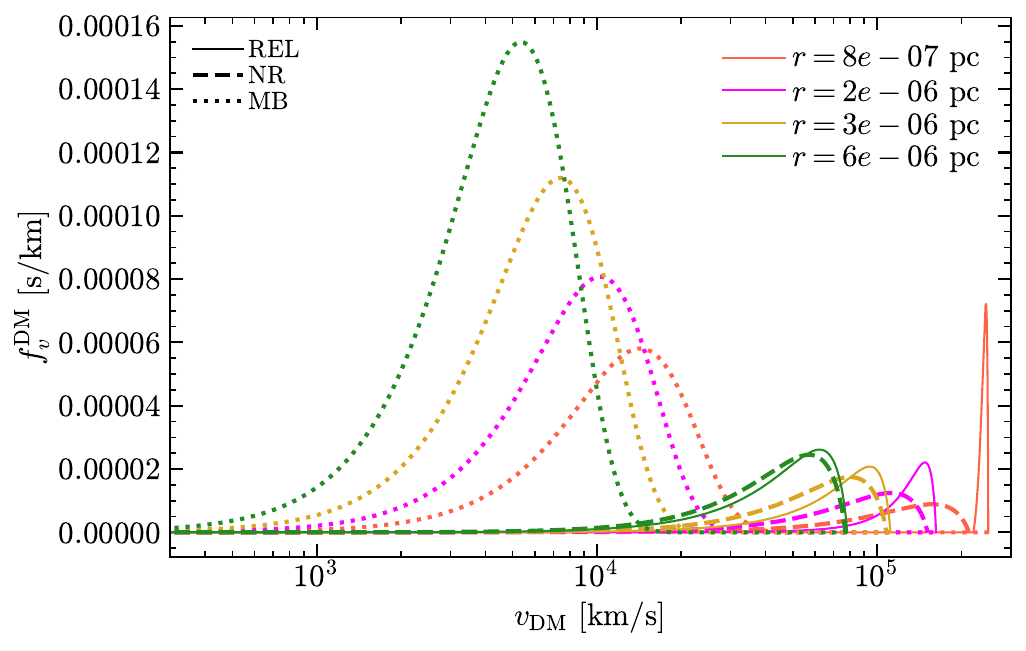}
    \caption{ Difference in VDF for DM reconstructed by the relativistic (\textit{REL}) and the non-relativistic (\textit{NR}) Eddington inversion, and the Maxwell - Boltzmann (\textit{MB}) distribution. We see examples for DM far (\textit{upper left}), close (\textit{upper right}), and very close (\textit{lower}) from the galactic center. Just the lower plot exhibits a visible different between the \textit{REL} and \textit{NR} approaches. }
    \label{fig:fv_distributions}
\end{figure*}
\begin{figure*}
    \centering
    \includegraphics[width=0.49\linewidth]{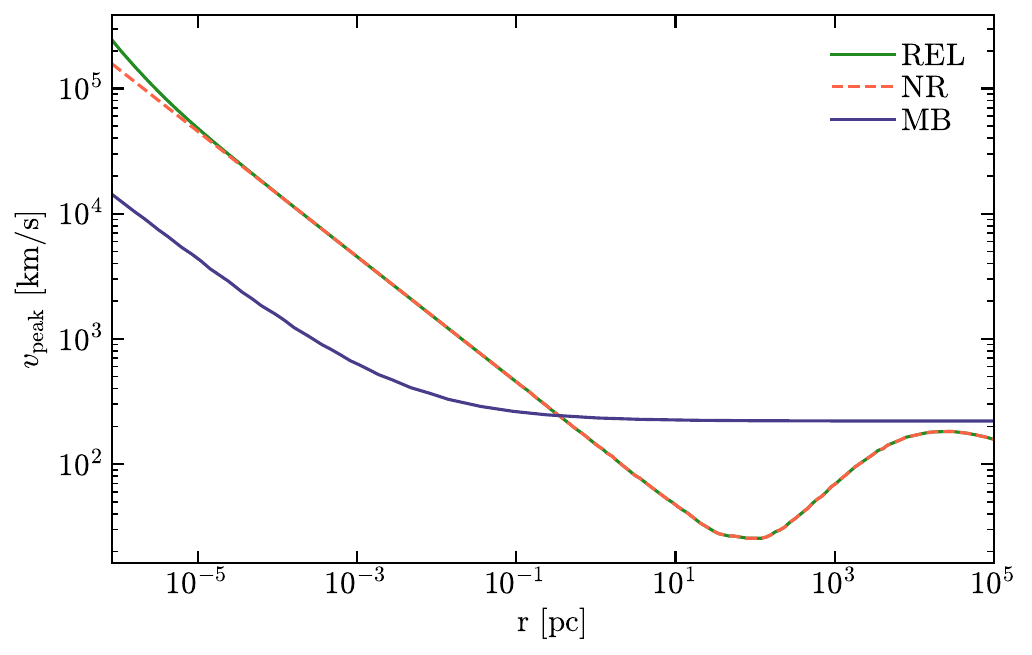}
    \includegraphics[width=0.49\linewidth]{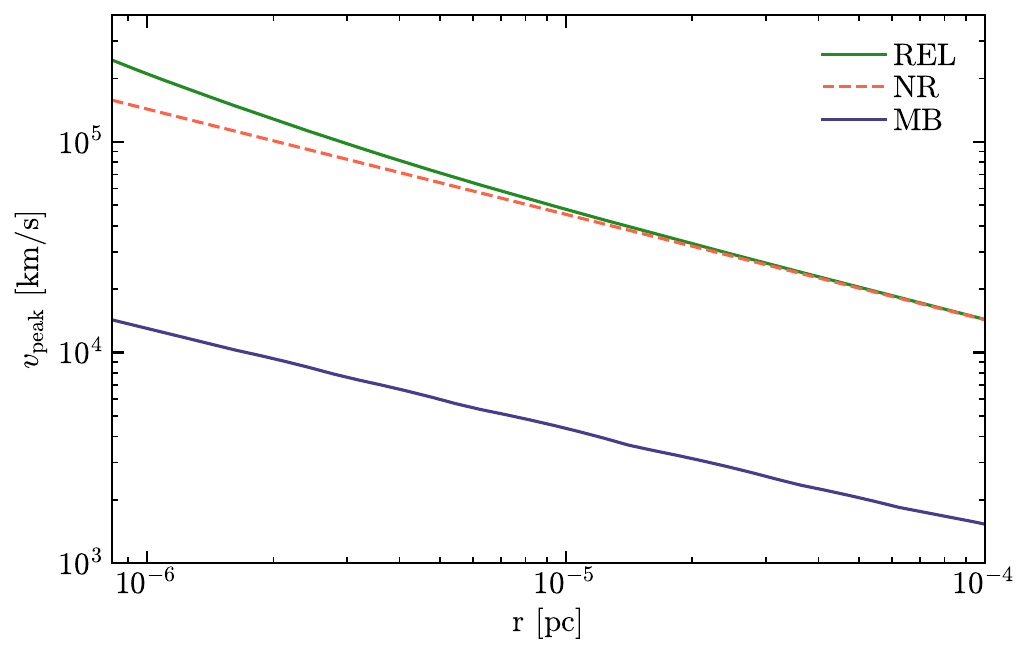}
    \includegraphics[width=0.49\linewidth]{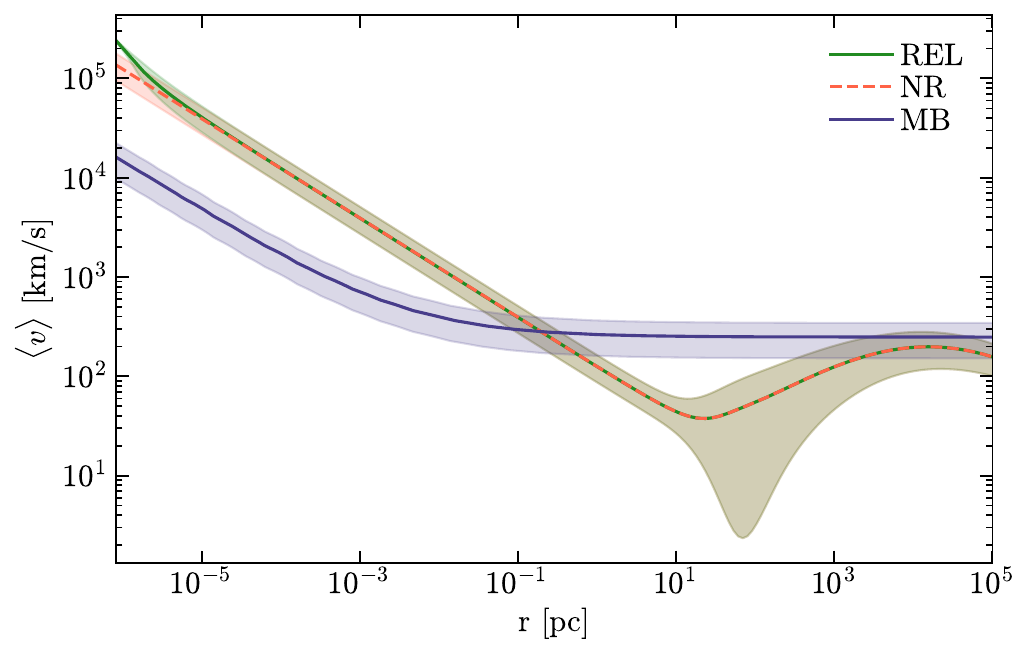}
    \includegraphics[width=0.49\linewidth]{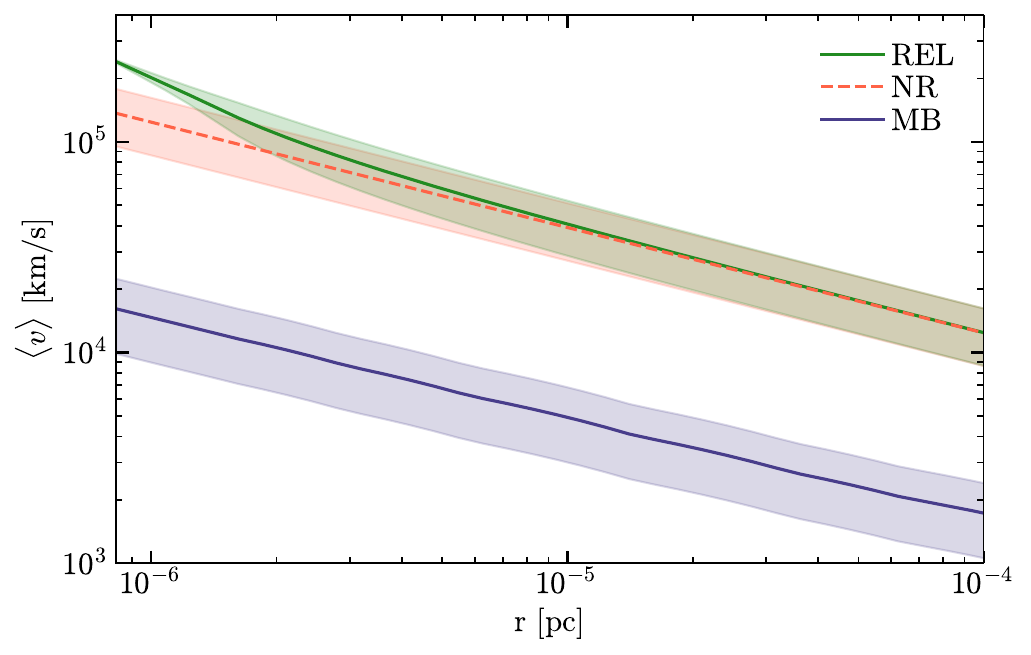}
    \caption{ Peak (\textit{upper plots}) and mean (\textit{lower plots}) velocities for the relativistic, non-relativistic (Eddington inversion) and Maxwell-Boltzmann distributions of DM. We can see the deviation of the relativistic computation close to the galactic center, on the \textit{left plots}. The mean velocity is shown together with a shadow covering the region $v_\mathrm{mean} \pm \sigma_v$, from the distributions.}
    \label{fig:v_peak_mean}
\end{figure*}

\subsection{Dark matter density profile in the Milky Way}

Regardless of the super-massive black hole (BH) at the center of our galaxy, there is a halo DM distribution which can be accounted for by the parametrization by Navarro - Frenk - White (NFW). After performing thorough simulations of the evolution of galaxies and observe how DM distributes at long times as a steady state, the NFW profile is postulated as an Ansatz, to describe what the simulations show. It takes the form~\cite{Navarro:1995iw}
\begin{equation}
\rho_\mathrm{NFW} (r) = \rho_\mathrm{crit} \frac{\delta_c}{(r / r_\mathrm{NFW}) \left( 1 + r/r_\mathrm{NFW} \right)^2}\,,
\end{equation}
where $r_\mathrm{NFW}$ is a scale radius, $\delta_c$ is a characteristic density, and $\rho_\mathrm{crit} \equiv 3H^2 / 8\pi G$ is the critical density for closure. This model was generalized to what we will cal the generalized NFW, or gNFW~\cite{Navarro:1996gj},
\begin{equation}
\label{eq:rho_gNFW}
\rho_\mathrm{gNFW} (r) = \rho_s \left( \frac{r_\mathrm{NFW}}{r} \right)^{\gamma_\mathrm{NFW}} \left( 1 + \frac{r}{r_\mathrm{NFW}} \right)^{\gamma_\mathrm{NFW} - 3}\,,
\end{equation}
with $\rho_s$ a scaled density and $\gamma_\mathrm{NFW}$ an index describing how the density scales with the radius. In the case of our Galaxy, standard values are $\rho_s \approx 0.0156 \, M_\odot / \mathrm{pc}^3$ and $r_\mathrm{NFW} = 1.5 \times 10^4$ pc~\cite{McMillan:2016jtx}. The value of $\gamma$ should be less than $0.83$, from S-stars constraints~\cite{Shen:2023kkm}. A standard value for $\gamma_\mathrm{NFW}$ is $0.8$~\cite{Yang:2024jtp}.

In the presence of Sagittarius A$^*$, our super-massive BH of $4.297 \times 10^{6} \, M_\odot$, DM is expected to accumulate in a very dense profile known as the \textit{DM spike}. Though a precise computation of the spike profile requires the computation of the distribution of the halo DM and the assumption of the conservation of the radial action, $I_r \equiv \oint v_r \, dr$, after the BH is formed, Silk and Gondolo~\cite{Gondolo_1999} parametrized the result, which for a gNFW halo distribution yields,
\begin{equation}
\label{eq:rho_sp}
\rho_\mathrm{sp} (r) = \rho_\mathrm{sp}^0 \times \left( \frac{r_\mathrm{sp}}{r} \right)^{\gamma_\mathrm{sp}}\,,
\end{equation}
where $\gamma_\mathrm{sp} = (9 - 2 \gamma_\mathrm{NFW}) / (4 - \gamma_\mathrm{NFW})$, $r_\mathrm{sp} = a_\gamma r_\mathrm{NFW} \left( \frac{M_\mathrm{BH}}{\rho_s r_\mathrm{NFW}^3} \right)^{1 / (3 - \gamma_\mathrm{NFW})}$. The quantity $a_\gamma$ depends on the slope of the gNFW profile, which for $\gamma_\mathrm{NFW} = 0.8$ is equal to $0.135$~\cite{Shen:2023kkm}. Finally, $\rho_\mathrm{sp}^0 = \rho_\mathrm{gNFW} (r = r_\mathrm{sp})$. This profile is valid for distances for the galactic center from $2 \, r_s$, where $r_s$ is the Schwarzschild radius of Sgr A$^*$, to $r_\mathrm{sp}$. Therefore, we can model the total DM density profile as,
\begin{equation}
\label{eq:rho_tot}
\rho_\mathrm{DM}(r) = \begin{cases}
    0 &\text{for } r < 2 \, r_s\,,\\
    \rho_\mathrm{sp} (r) &\text{for } 2 \, r_s \leq r < r_\mathrm{sp} \,,\\
    \rho_\mathrm{gNFW} (r) &\text{for } r_\mathrm{sp} \leq r \,.
\end{cases}
\end{equation}
\begin{figure*}
    \centering
    \includegraphics[width=0.49\linewidth]{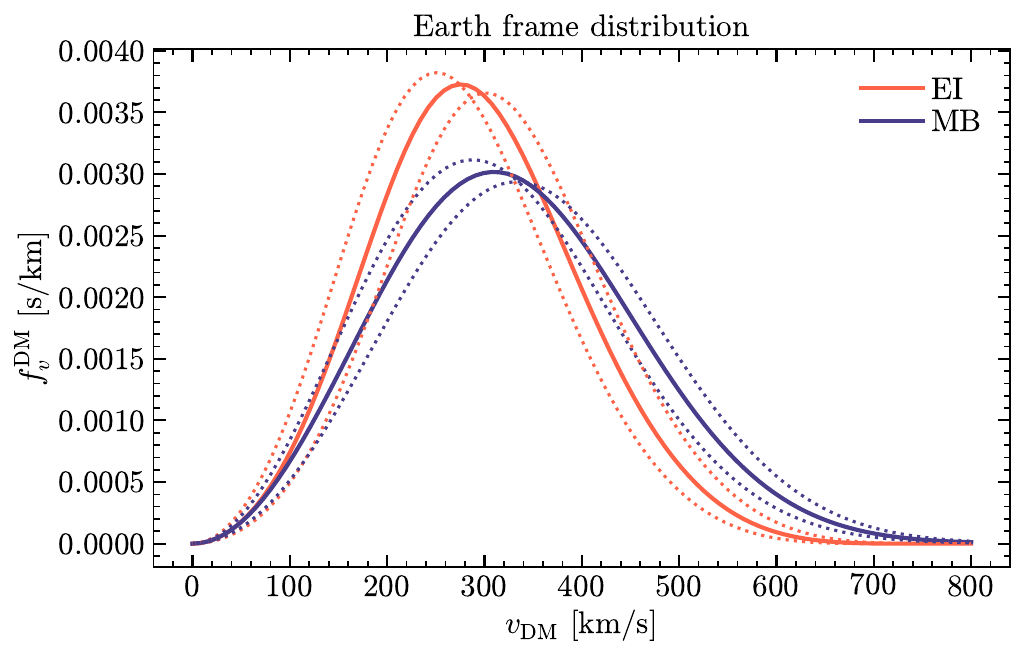}
    \includegraphics[width=0.49\linewidth]{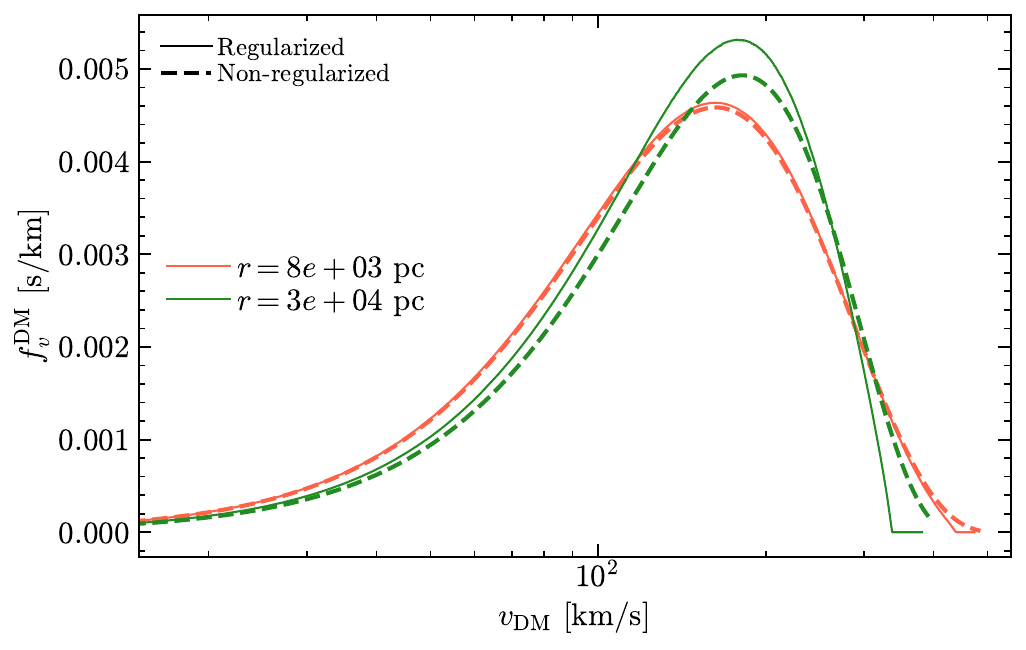}
    \caption{ Velocity distribution for local DM and DM far from the center of the galaxy. The plot on the \textit{left} shows the distributions after Eddington inversion (EI) and Maxwell-Boltzmann (MB) as seen from Earth. The dotted lines show the range of distributions when the Earth is moving anti-parallel or parallel to the Sun. On the \textit{right}, we see the effect of regularizing the relativistic Eddington inversion with respect to dropping the boundary term. For this non-relativistic regimes, we do not distinguish between non-relativistic and relativistic Eddington inversion, since they yield the same results.}
    \label{fig:fv_local}
\end{figure*}

It is important to mention that when DM particles can annihilate, then the spike is modified. We will assume there is no annihilation that visibly alters the profile.

\subsection{Velocity distribution of dark matter}

With the tools presented in the previous sections, we can now compare the VDFs we obtain from the different approaches: (1) the \textit{relativistic} (REL) one refers to the fully relativistic Eddington inversion, the method developed in the present work in \Cref{sec:GREddInv}; (2) the \textit{non relativistic} (NR) refers to the standard Eddington inversion, as shown in \Cref{sec:EddInv}; (3) the \textit{Maxwell - Boltzmann} (MB) regards the MB velocity distribution, with the dispersion found as explained in \Cref{sec:MB}. We regard DM to be distributed following the Gondolo - Silk spike with a NFW halo. We will neglect the effect of baryonic matter. 

In \Cref{fig:fv_distributions} we can see three different plots for the three distributions for different regions in the Milky Way. The top left plot shows distributions for regions not close from the galactic center: from $10$ pc to $30$ kpc. We see that REL and NR yield both the same result, while MB is more invariant with higher predicted velocities. The top right plot consider distances close to the galactic center, from $10^{-5}$ to $4 \times 10^{-2}$ pc. For this region, REL and NR are still sufficiently close, while MB predicts lower velocities. Finally, the lower plot goes to the region closest to Sgr A$^*$: from $2 \, r_s$ to $\sim 6 \times 10^{-6}$ pc. Here MB predicts even lower values of velocity, while REL starts to deviate from NR, yielding higher relativistic velocities and with relatively less spread.

These results can also be schematically seen by following the peak velocity of the distribution $v_\mathrm{peak}$ or its mean value $\langle v \rangle$, shown in \Cref{fig:v_peak_mean}. The top plots show the peak velocity for the whole galaxy (left) and for the region closest to Sgr A$^*$ (right). At around $1$ pc the three distributions coincide. However, for larger values of $r$, MB gives a larger result, while smaller for positions closer to the center of the galaxy. As we approach Sgr A$^*$, the velocities grow dramatically. The MB starts growing at $\mathcal{O} \left(10^{-2} \right)$ pc, while the other at $\mathcal{O} \left(10^{2} \right)$ pc. As we approach to $\mathcal{O} \left(10^{-5} \right)$ pc, REL deviates from NR and exhibits a higher peak velocity up to $\sim 2$ times the NR as we approach to $r = 2 \, r_s$. A very similar behavior can be seen for the mean velocity in the lower plots. We are also showing a shaded area containing $\langle v \rangle \pm \sigma_v$. We see that as we approach to the very center of the galaxy, REL produces a much more concentrated distribution, with visibly less spread.

In the left plot of \Cref{fig:fv_local}, we also show the local DM distributions as seen from the Earth. Here \textit{EI} (Eddington Inversion) refers to REL or NR, since they yield the same result. We are transforming the distribution far from Earth, to the one as seen from Earth, since we are moving with respect to DM. Our motion is described by,
\begin{equation}
\vec{v}_\mathrm{obs} = \vec{v}_\odot + \vec{v}_\oplus\,,
\end{equation}
where $\vec{v}_\mathrm{obs}$ is the velocity of our observation with respect to DM, $\vec{v}_\odot$ is the velocity of the Sun with respect to DM and $\vec{v}_\oplus$ is the velocity of Earth with respect to the Sun. The speed at which we observe DM is then bounded: $v_\odot - v_\oplus \le v_\mathrm{obs} \le v_\odot + v_\oplus$. The solid lines of the plot are showing $v_\mathrm{obs} = v_\odot$, while the dotted lines show $v_\mathrm{obs} = v_\odot \pm v_\oplus$. Eddington inversion predicts lower velocities and a reduced high velocity tail.

In the right plot of \Cref{fig:fv_local}, we are showing the results by REL locally ($8$ kpc) and farther away ($30$ kpc) for the non-regularized Eddington inversion (dashed line), so just dropping the boundary term, and the regularized one (solid line), following the procedure explained in \Cref{sec:regularization}. The effect is more visible as we go to the boundary of the galaxy, where there is a sharper cut on the high velocity tail, while the peak is slightly reduced. This is produced by the transformation of the density on the boundary, where the impact fo the transformation $\rho \to \tilde{\rho}$ is more visible.

\subsection{Accretion rate of dark matter into neutron stars}

The differences we have found in velocities can have a visible impact when we estimate interactions in which DM is involved. To have an idea of how these estimations change, we will compute a factor that is present in accretion rates of DM into neutron stars (NS). Following the same approach as in \cite{Bell:2020jou, Davoudiasl:2025vpe}, we can compute how much DM is captured by a NS per time, as seen from far from far from the star. If the velocities of DM far from the star are much smaller than the boost given by the NS, which is even true for the highest velocities we are dealing with, then the capture rate takes the following form,
\begin{equation}
\begin{split}
C = 4 \pi A_\mathrm{acc} \frac{\rho}{m} \int dr \, r^2 \sqrt{g_{rr}} \frac{\sqrt{1 - g_{tt}}}{g_{tt}} \times \Omega^-_v (r)\,,
\end{split}
\end{equation}
where $\rho$ and $m$ are the density and mass of DM respectively, $g_{rr}$ and $g_{tt}$ are the positive metric elements produced by the star, $\Omega^-_v$ is the probability rate for DM of interacting with the stellar material and become bound to the star. Since the curvature produced by Sgr A$^*$ varies spatially much less than the one produced by the NS, we can regard the space on which the star is moving during the accretion as flat, so that the metric elements are just produced by the NS. In that case, there is an extra factor needed, in order to account for the gravitational redshift from the field produced by the super-massive BH. Finally, $A_\mathrm{acc}$ is a factor that comes from integrating over the DM distribution. More details about its computation can be found in the Appendix. Under a classical approach, this factor is,
\begin{equation}
\begin{split}
A^\mathrm{NR}_\mathrm{acc} = &\frac{1}{2} \int dv \, v \\&\times\int d\cos \theta \frac{f_\mathrm{DM} \left( \sqrt{v^2 + v_\mathrm{NS}^2 + 2 \, v \, v_\mathrm{NS} \cos \theta } \right)}{v^2 + v_\mathrm{NS}^2 + 2 \, v \, v_\mathrm{NS} \cos \theta}\,,
\end{split}
\end{equation}
where $v_\mathrm{NS}$ is the speed of the NS with respect to DM. The expression is independent of the direction of motion, so that DM may interact not isotropically in the star, but the expression is just dependent on the radial position. Therefore, the angle between the star's motion and the DM particle involved, $\theta$, is integrated out without loss of generality.

A relativistic expression for the factor is,
\begin{equation}
\begin{split}
A^\mathrm{REL}_\mathrm{acc} = \frac{1}{2} \int dv \, &v\int d\cos \theta \, \sqrt{\frac{1 - \tilde{v}^2}{1 - v^2}} \\&\times \frac{(1 - v_\mathrm{NS}^2) (v + v_\mathrm{NS} \cos \theta) }{ \tilde{v}^3\left( 1 + v \, v_\mathrm{NS} \cos \theta \right)^3} f_\mathrm{DM} (\tilde{v})\,,
\end{split}
\end{equation}
where $\tilde{v}$ should be expressed in terms of $v$ and $\cos \theta$ as in \Cref{eq:vtilde_REL}.

 In the case of the standard Maxwell - Boltzmann velocity distribution, the factor takes the form,
\begin{equation}
\begin{split}
A_\mathrm{acc}^\mathrm{MB} = \frac{1}{v_\mathrm{NS}} \times \mathrm{erf} \left( \sqrt{\frac{3}{2}} \frac{v_\mathrm{NS}}{v_d} \right)\,,
\end{split}
\end{equation}

\begin{figure*}
    \centering
    \includegraphics[width=0.49\linewidth]{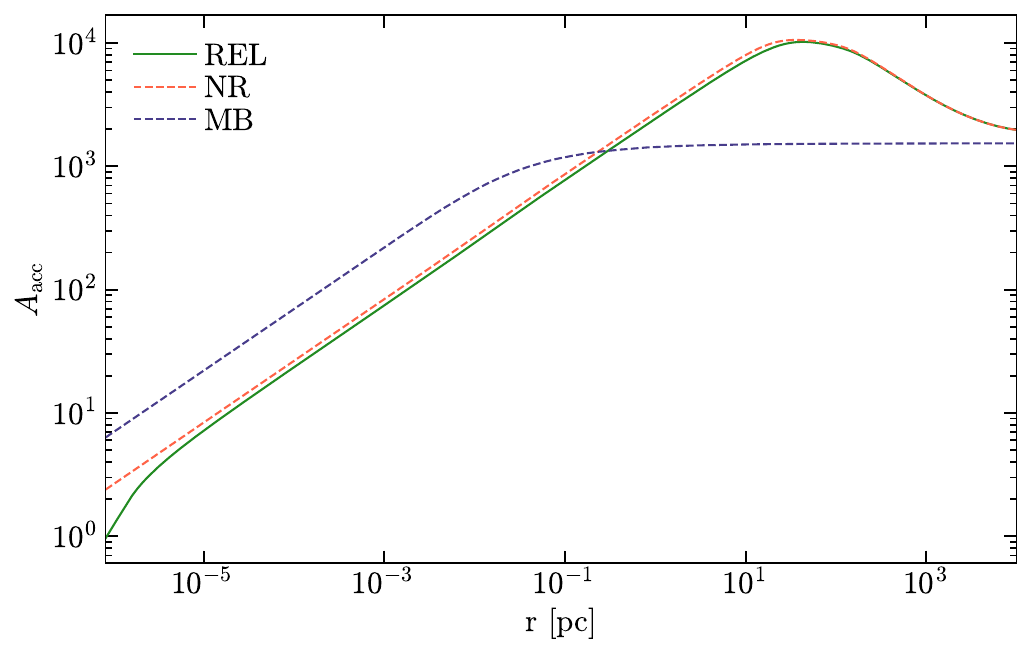}
    \includegraphics[width=0.49\linewidth]{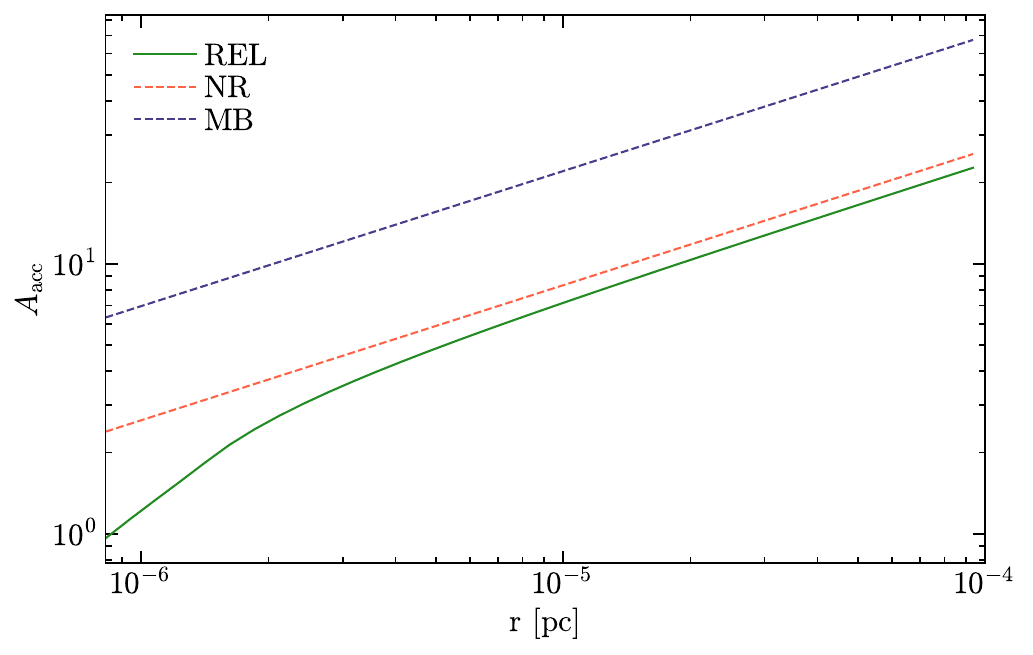}
    \caption{ Accretion factors used for computation of capture rates of DM in neutron stars, computed under the three different approaches for the PSDF reconstruction: relativistic, non-relativistic and Maxwell-Boltzmann. On the \textit{right plot} we can see the behavior very close to Sgr A$^*$. }
    \label{fig:accretion}
\end{figure*}

We have computed the accretion factor for the three distributions, as can be seen in \Cref{fig:accretion}. As we can see from the plot, the factor exhibits the opposite behavior as velocity: the higher the velocities, the smaller the amount of DM that falls into the star. As expected from the previous results, NR and REL yield the closer results for $r \gtrsim 10^{-5}$ pc. However, they still differ up to $\sim 10$ pc, because there is a relativistic correction also coming from the definition of the VDF from the star frame. For large distances, REL and NR have larger accretion rates than MB, while lower close to the galactic center. Very close to Sgr A$^*$, REL produce an accretion factor almost one order lower than the MB one. Therefore, computing accretion in this region with the MB distribution may overestimate results by one order. If computations are done by the classical Eddington inversion, we may overestimate results by a factor of $\sim 2$.

%

%%%%%%%%%%%%%%%%%%%%%%%%%%%%%%%
%        SUMMARY
\section{Summary}
\label{sec:Conc}

In the present work, we have developed a method to infer the phase space distribution function of dark matter, together with its velocity distribution, from the density profile of dark matter. We have departed from the assumption that dark matter behaves like a perfect fluid in a static metric and we have assumed the Tolman-Oppenheimer-Volkoff equations to reconstruct the pressure and the metric elements. After some algebraic manipulations, we have used the relation between the density and the temporal metric element, $g_{tt} \equiv |g_{tt}|$, to establish an equality between a derivative expression of the density and $g_{tt}$ that takes the form of an Abel-type integral equation. This could be inverted using an Eddington inversion-like approach that fully respects General Relativity. The solution allows us to reconstruct the dark matter distribution as a function of the energy per mass of each particle as seen from infinity, $\mathcal{E}$. This distribution also allowed us to compute the dark matter velocity distribution, under the assumption of being located in a particular radial position in the galaxy.

We have shown the results for the previous computations for the Milky Way, by assuming a Navarro - Frenk - White halo plus a Gondolo - Silk dark matter spike close to the galactic center. We have tested our relativistic approach, which reproduces the standard Eddington inversion approach and takes its for under the classical approximation. These results start diverging from each other at the order of $\mathcal{O}(10^{-5})$ pc, towards Sgr A$^*$. In order to compute the distributions under the classical Eddington inversion method, we have calculated analytical expressions for the gravitational potential and derivatives of the density profile with respect to it, under the Milky Way model used. The numerical instability of the classical method for regions close to the galactic center, forced us to take this approach. However, the relativistic method is sufficiently stable at any position in the Milky Way, so no analytical expressions were needed. In all the computations, we have neglected the effect of baryonic matter.

We have also computed Maxwell - Boltzmann velocity distributions for dark matter, the standard method for direct detection experiments. We have found that inversion methods yield visible differences with this standard approach. 

Finally, these differences among methods are detectable when we compute the accretion rate factor of dark matter into stars, and particularly into neutron stars, as shown at the end of the previous section. We have derived a general formula for capture rates of dark matter in neutron stars given a general velocity distribution far from the star under a classical and also a relativistic approach. We have shown how each of the three methods yield different results. Given the precision era experienced in particle and astroparticle physics, more thorough and precise methods are certainly needed.

\begin{acknowledgments}
\titleprl{Acknowledgments}
J.~H.~Z. is supported by the National Science Centre, Poland (research grant No. 2021/42/E/ST2/00031). The research of PP is funded by Tamkeen under the grant ADHPG- AD457 to NYUAD.

\end{acknowledgments}

% =========================
% Appendix
% =========================
\clearpage
\onecolumngrid
\appendix

%%%%%%%%%%%%%%%%%%%%%%%%%%%%%%%
%   ACCRETION FACTOR
\section{Accretion factor for a generalized dark matter distribution}

In order to compute the accretion factor, we need to compute the capture rate of DM by a NS. We will follow the standard procedure established by \cite{Press:1985ug} and \cite{Gould:1987ir}, and then worked out for NSs by \cite{Bell:2020jou}.

\subsection{Distribution as seen from the NS: classical approach}

The main aim is to transform the distribution to the frame of a moving star with velocity $\vec{v}_\mathrm{NS}$. Since we are not interested in effects arising form DM particles non isotropically distributed within the star, since interactions are only dependent on energies, which follow directly from the radial position, we may integrate out any angular dependence. We will use tilde variables for the DM frame to distinguish them from the NS frame.

Far from the NS, DM has a velocity distribution $f_\mathrm{DM} (v)$, such that $f_\mathrm{DM} (\tilde{v}) \equiv 4 \pi \tilde{v}^2 f_{\mathrm{DM}, \vec{v}} (\vec{\tilde{v}})$, as follows from its isotropic distribution. The relation with the 3D-velocity distribution is crucial, since as we move towards the NS frame, the new distributions are related by,
\begin{equation}
\label{eq:rel_dists}
f_{\mathrm{DM}, \vec{v}} \, d^3 \tilde{v} = f_{\mathrm{NS}, \vec{v}} \, d^3 v\,.
\end{equation}

Since the transformation between velocities is just a translation, the Jacobian is equal to one. The relation between the velocities is,
\begin{equation}
\tilde{v}^2 = v^2 + v_\mathrm{NS}^2 + 2 \, v_\mathrm{NS} \, v \cos \theta\,,
\end{equation}
where $\theta$ is the angle between $\vec{v}_\mathrm{NS}$ and $\vec{v}$. Therefore, by following the equality in \Cref{eq:rel_dists}, we find,
\begin{equation}
\label{eq:rel_dists}
f_{\mathrm{DM}, \vec{v}} \, d^3 \tilde{v} = f_{\mathrm{NS}, \vec{v}} \, d^3 v = f_{\mathrm{DM}, \vec{v}} \left( \sqrt{v^2 + v_\mathrm{NS}^2 + 2 \, v_\mathrm{NS} \, v \cos \theta} \right) d^3 v \,,
\end{equation}
where we used $d^3 \tilde{v} = d^3 v$.  Then, we do $d^3 v \equiv v^2 \, dv \, d\cos \theta \, d \phi$ and take into account that $f_\mathrm{DM} (\tilde{v}) = 4 \pi \tilde{v}^2 f_{\mathrm{DM}, \vec{v}} (\tilde{v})$. Then, the 1D-velocity distribution is given by,
\begin{equation}
\label{eq:fv_NS}
f_\mathrm{NS} (v) = \frac{v^2}{2} \int  \frac{f_\mathrm{DM} \left( \sqrt{v^2 + v_\mathrm{NS}^2 + 2 \, v_\mathrm{NS} \, v \cos \theta} \right)}{v^2 + v_\mathrm{NS}^2 + 2 \, v_\mathrm{NS} \, v \cos \theta} \, d\cos \theta\,.
\end{equation}

\subsection{Distribution as seen from the NS: relativistic approach}

In the relativistic case, we need to consider the following equality that relates the different distributions,
\begin{equation}
\label{eq:rel_dists_REL}
f_{\mathrm{DM}, \vec{v}} \, d^3 \tilde{v} \, d^3 \tilde{r} = f_{\mathrm{NS}, \vec{v}} \, d^3 v \, d^3 r\,,
\end{equation}
which follow from the Lorentz invariance of $d^3 p \, d^3 r$. This entails:
\begin{equation}
\label{eq:rel_dists_REL_final}
f_{\mathrm{NS}, \vec{v}} (v) = J \left( \frac{\tilde{v}, \tilde{r}}{v, r} \right) \, f_{\mathrm{DM}, \vec{v}} (\tilde{v})\,,
\end{equation}
where $J$ is the Jacobian of the transformation. We can express $\tilde{v}^2$ as,
\begin{equation}
\label{eq:vtilde_REL}
\tilde{v}^2 = \frac{(v \cos \theta + v_\mathrm{NS})^2 + v^2 \sin^2 \theta (1 - v_\mathrm{NS}^2)}{\left( 1 + v \, v_\mathrm{NS} \cos \theta \right)^2}\,.
\end{equation}

With this relation, we can find the Jacobian, which can be decomposed into two factors: $J(\tilde{r} / r) = \sqrt{(1 - \tilde{v}^2) / (1 - v^2)}$ and
\begin{equation}
\frac{d\tilde{v}}{dv} = \frac{(1 - v_\mathrm{NS}^2) (v + v_\mathrm{NS} \cos \theta) }{ \tilde{v} \left( 1 + v \, v_\mathrm{NS} \cos \theta \right)^3}\,.
\end{equation}

Putting everything together, the 1D-velocity distribution as seen from the star is
\begin{equation}
\label{eq:fv_NS_REL}
f_\mathrm{NS} (v) = \frac{v^2}{2}\int d\cos \theta \, \sqrt{\frac{1 - \tilde{v}^2}{1 - v^2}} \frac{(1 - v_\mathrm{NS}^2) (v + v_\mathrm{NS} \cos \theta) }{ \tilde{v}^3\left( 1 + v \, v_\mathrm{NS} \cos \theta \right)^3} f_\mathrm{DM} (\tilde{v})\,,
\end{equation}
where $\tilde{v}$ should be put in terms of $v$ and $\cos \theta$, as in \Cref{eq:vtilde_REL}.

\subsection{Differential number density of DM in the star}

Now we need to know how much DM traversing the star there is at any radial position. We first need to compute the differential flux into the star putting ourselves far from the star, on a spherical shell of radius $R \to \infty$. This was originally found by \cite{Press:1985ug, Gould:1987ir},
\begin{equation}
d\Phi = \frac{1}{4} \frac{\rho}{m} f_\mathrm{NS}(v) \, v \, dv \, d \cos^2 \theta\,,
\end{equation}
where $\rho$ is the DM density, $m$ is its mass, and $\theta$ represents the angle between a DM particle very far from the star and a vector pointing from the DM particle to the center of the star.

We can re-parametrize this equality in terms of the angular momentum per mass ($L$). We also multiply it by $4 \pi R^2$ to get the number of particles per time passing through the shell, for velocities within $[v, v + dv]$ and momenta within $[L, L + dL]$,
\begin{equation}
d\dot{N} = 2 \pi \frac{\rho}{m} \frac{f_\mathrm{NS}(v)}{v} \, dv \, L \, dL\,.
\end{equation}

Now we can measure the number density. It is important to remember that, since we have integrated out the angular dependence of the fluxes, the result we will measure is the mean number density. We will let the flux move towards the star a time $dt$ so that we will generate a volume that has passed through the spherical shell at $R$. Then, we will compute the volume occupied by that initial number of particles when they get to a radius $r$ inside the star. This is precisely of compressed spherical shell with a volume $V_r = 4 \pi r^2 g_{rr} \, dr$. Therefore, the differential mean number density of DM at $r$, $d\bar{n}$, is,
\begin{equation}
\begin{split}
d\bar{n} = & \, 2 \pi \frac{\rho}{m} \frac{f_\mathrm{NS}(v)}{v} \, dv \, L \, dL \times \frac{dt}{4 \pi r^2 \sqrt{g_{rr}} \, dr}\,,\\
= & \, 2 \pi \frac{\rho}{m} \frac{f_\mathrm{NS}(v)}{v} \, dv \, L \, dL \times \frac{1}{4 \pi r^2 \sqrt{g_{tt}} \sqrt{1 / g_{tt} - 1 - L^2 / r^2}   }\,,
\end{split}
\end{equation}
where we have used the geodesic equations for a metric of the form $ds^2 = - g_{tt} dt^2 + g_{rr} dr^2 + r^2 d\Omega^2$, which accounts for the gravitational field in a NS. We are also considering that the boost of DM particles far from the star are negligible with respect to the energies at which the NS accelerates them inside the star.

\subsection{Capture rate}

The differential capture rate of DM in a NS is defined as,
\begin{equation}
\label{eq:CR_def}
dC = n \times \Omega^- \times dV\,,
\end{equation}
where $dV$ is a differential volume in the star and $\Omega^-$ is the probability rate, i.e., the probability of a particle of being captured per time, as seen far from the star. This is a well known quantity in the literature. We refer the reader to \cite{Bell:2020jou} for more details.

To obtain the total number density we need to integrate on L, from $0$ to $\sqrt{1 / g_{tt} - 1} \times r$, which are the allowed values at $r$, and $v$.
\begin{equation}
\begin{split}
\bar{n} = \frac{\rho}{m} \left( \int \frac{f_\mathrm{NS}(v)}{v} \, dv \right) \frac{\sqrt{1 - g_{tt}}}{\left(g_{tt} \right)^{3/2}}\,,
\end{split}
\end{equation}
where we have included a factor of $2$ to account for incoming and outgoing fluxes.

Finally, we use \Cref{eq:CR_def} to find the total capture rate. We also need to multiply by $\sqrt{g_{tt}}$ to account for the time being measured far from the star.
\begin{equation}
\begin{split}
C = 4 \pi \frac{\rho}{m} \left( \int \frac{f_\mathrm{NS}(v)}{v} \, dv \right) \int dr \, r^2 \sqrt{g_{rr}} \frac{\sqrt{1 - g_{tt}}}{g_{tt}} \times \Omega^- (r)\,.
\end{split}
\end{equation}

The classical accretion factor would then be,
\begin{equation}
\label{eq:A_acc_NR}
A^\mathrm{NR}_\mathrm{acc} = \frac{1}{2} \int dv \, v \int d\cos \theta \frac{f_\mathrm{DM} \left( \sqrt{v^2 + v_\mathrm{NS}^2 + 2 \, v_\mathrm{NS} \, v \cos \theta} \right)}{v^2 + v_\mathrm{NS}^2 + 2 \, v_\mathrm{NS} \, v \cos \theta}\,,
\end{equation}
while the relativistic one is,
\begin{equation}
\label{eq:A_acc_REL}
A^\mathrm{REL}_\mathrm{acc} = \frac{1}{2} \int dv \, v\int d\cos \theta \, \sqrt{\frac{1 - \tilde{v}^2}{1 - v^2}} \frac{(1 - v_\mathrm{NS}^2) (v + v_\mathrm{NS} \cos \theta) }{ \tilde{v}^3\left( 1 + v \, v_\mathrm{NS} \cos \theta \right)^3} f_\mathrm{DM} (\tilde{v})\,,
\end{equation}
where $\tilde{v}$ should be replaced as in \Cref{eq:vtilde_REL}.

%%%%%%%%%%%%%%%%%%%%%%%%%%%%%%%
%   SOME ANALYTIC EXPRESSIONS
\section{Analytic expressions for the non-relativistic case}

The non-relativistic Eddington inversion relies on the computation of three main quantities: the potential $\Psi \equiv \Phi_0 - \Phi$ and the derivatives $d\rho / d\Psi$ and $d^2\rho / d\Psi^2$. These quantities can be computed semi-analytically for some density profiles.

\subsection{Gondolo Silk + gNFW}

If we consider the Gondolo Silk spike plus a gNFW halo for DM and we disregard the effect of any other source of matter, but just from DM and Sgr A$^*$, then we can have a full semi-analytical description of the quantities needed for the non-relativistic Eddington inversion. This is particularly important to obtain non-noisy results for Eddington inversion close to the center of the galaxy. 

All the quantities have different explicit expressions for two regions: $2 \, r_s \leq r < r_\mathrm{sp}$ (region A) and $r_\mathrm{sp} \leq r$ (region B). We will use these symbols as subscripts to distinguish between the regions.

The accumulated mass takes the form,
\begin{equation}
\begin{split}
M_\mathrm{A} (r) = &M_{\text{BH}}+\frac{4 \pi  \rho _{\text{sp}} r_{\text{sp}}^{\gamma _{\text{sp}}} \left(2^{3-\gamma _{\text{sp}}} r_s^{3-\gamma _{\text{sp}}}-r^{3-\gamma
   _{\text{sp}}}\right)}{\gamma _{\text{sp}}-3}\,,\\
M_\mathrm{B} (r) = &M_{\text{BH}}-4 \pi  (-1)^{\gamma _{\text{NFW}}} r_{\text{NFW}}^3 \rho _s \left(B_{-\frac{r}{r_{\text{NFW}}}}\left(3-\gamma _{\text{NFW}},\gamma
   _{\text{NFW}}-2\right)-B_{-\frac{r_{\text{sp}}}{r_{\text{NFW}}}}\left(3-\gamma _{\text{NFW}},\gamma _{\text{NFW}}-2\right)\right) \\&+\frac{\pi  \left(2^{5-\gamma
   _{\text{sp}}} \rho _{\text{sp}} r_{\text{sp}}^{\gamma _{\text{sp}}} r_s^{3-\gamma _{\text{sp}}}-4 r_{\text{sp}}^3 \rho _{\text{sp}}\right)}{\gamma _{\text{sp}}-3}\,,
\end{split}
\end{equation}
where the function $B$ is the analytically continued unregularized incomplete beta function, defined as,
\begin{equation}
\begin{split}
B_z (a, b) = \int_0^z t^{a - 1} (1 - t)^{b - 1} \, dt\,,
\end{split}
\end{equation}
where this function is analytically continued when $z < 0$. With this expression, we can compute $\Psi$:
\begin{equation}
\begin{split}
\Psi_\mathrm{A} (r) = &G \Bigg[M_{\text{BH}} \left(\frac{1}{r}-\frac{1}{r_{\max }}\right)+4 \pi  \bigg(-\frac{r_{\text{NFW}}^2 \rho _s \left(\frac{r_{\max }}{r_{\max
   }+r_{\text{NFW}}}\right){}^{2-\gamma _{\text{NFW}}}}{\gamma _{\text{NFW}}-2}+\frac{\left(r_{\max }-r\right) 2^{3-\gamma _{\text{sp}}} \rho _{\text{sp}}
   r_{\text{sp}}^{\gamma _{\text{sp}}} r_s^{3-\gamma _{\text{sp}}}}{r r_{\max } \left(\gamma _{\text{sp}}-3\right)} \\&+\frac{r_{\text{sp}}^3 \rho _{\text{sp}}}{r_{\max }
   \left(\gamma _{\text{sp}}-3\right)}+\frac{r_{\text{NFW}}^2 \rho _s \left(\frac{r_{\text{sp}}}{r_{\text{NFW}}+r_{\text{sp}}}\right){}^{2-\gamma _{\text{NFW}}}}{\gamma
   _{\text{NFW}}-2}-\frac{\rho _{\text{sp}} r^{2-\gamma _{\text{sp}}} r_{\text{sp}}^{\gamma _{\text{sp}}}}{\gamma _{\text{sp}}^2-5 \gamma
   _{\text{sp}}+6}-\frac{r_{\text{sp}}^2 \rho _{\text{sp}}}{\gamma _{\text{sp}}-2}\bigg) \\&+\frac{4 \pi  (-1)^{\gamma _{\text{NFW}}} r_{\text{NFW}}^3 \rho _s
   \left(B_{-\frac{r_{\max }}{r_{\text{NFW}}}}\left(3-\gamma _{\text{NFW}},\gamma _{\text{NFW}}-2\right)-B_{-\frac{r_{\text{sp}}}{r_{\text{NFW}}}}\left(3-\gamma
   _{\text{NFW}},\gamma _{\text{NFW}}-2\right)\right)}{r_{\max }}\Bigg]\,,\\
\Psi_\mathrm{B} (r) = &G \Bigg[M_{\text{BH}} \left(\frac{1}{r}-\frac{1}{r_{\max }}\right)-4 \pi  (-1)^{\gamma _{\text{NFW}}} r_{\text{NFW}}^3 \rho _s \Bigg(\frac{\left(-\frac{r_{\max
   }}{r_{\max }+r_{\text{NFW}}}\right){}^{2-\gamma _{\text{NFW}}}-\left(-\frac{r}{r_{\text{NFW}}+r}\right){}^{2-\gamma _{\text{NFW}}}}{\left(\gamma _{\text{NFW}}-2\right)
   r_{\text{NFW}}} \\&-\frac{B_{-\frac{r_{\max }}{r_{\text{NFW}}}}\left(3-\gamma _{\text{NFW}},\gamma _{\text{NFW}}-2\right)}{r_{\max
   }}+\frac{B_{-\frac{r}{r_{\text{NFW}}}}\left(3-\gamma _{\text{NFW}},\gamma _{\text{NFW}}-2\right)}{r}\Bigg) \\&+4 \pi  \left(\frac{1}{r}-\frac{1}{r_{\max }}\right)
   (-1)^{\gamma _{\text{NFW}}} r_{\text{NFW}}^3 \rho _s B_{-\frac{r_{\text{sp}}}{r_{\text{NFW}}}}\left(3-\gamma _{\text{NFW}},\gamma _{\text{NFW}}-2\right) \\&+\frac{\pi 
   \left(\frac{1}{r}-\frac{1}{r_{\max }}\right) 2^{5-\gamma _{\text{sp}}} \rho _{\text{sp}} r_{\text{sp}}^{\gamma _{\text{sp}}} r_s^{3-\gamma _{\text{sp}}}}{\gamma
   _{\text{sp}}-3}-\frac{4 \pi  \left(\frac{1}{r}-\frac{1}{r_{\max }}\right) r_{\text{sp}}^3 \rho _{\text{sp}}}{\gamma _{\text{sp}}-3}\Bigg]\,,
\end{split}
\end{equation}
where $G$ is Newton's gravitational constant. Although the explicit complex factors that the formula exhibits, the overall result is real. 

With the help of the expression for the density and for the potential, we can construct the derivatives in the following way:
\begin{equation}
\begin{split}
\frac{d \rho}{d \Psi} &= \frac{d\rho / dr}{ d\Psi / dr}\,,\\
\frac{d^2 \rho}{d \Psi^2} &= \frac{\left( d^2 \rho / dr^2 \right) d\Psi / dr - \left( d^2 \Psi / dr^2 \right) d\rho / dr}{\left( d\Psi / dr \right)^3}\,.
\end{split}
\end{equation}

With this expressions, we get:
\begin{equation}
\begin{split}
\frac{d \rho}{d \Psi}_\mathrm{A} (r) = &\frac{r 2^{\gamma _{\text{sp}}} \left(\gamma _{\text{sp}}-3\right) \gamma _{\text{sp}} \rho _{\text{sp}} \left(r_s r_{\text{sp}}\right){}^{\gamma _{\text{sp}}}}{G
   \left(M_{\text{BH}} 2^{\gamma _{\text{sp}}} \left(\gamma _{\text{sp}}-3\right) \left(r r_s\right){}^{\gamma _{\text{sp}}}+4 \pi  \rho _{\text{sp}} \left(8 r_s^3
   \left(r r_{\text{sp}}\right){}^{\gamma _{\text{sp}}}-r^3 2^{\gamma _{\text{sp}}} \left(r_s r_{\text{sp}}\right){}^{\gamma _{\text{sp}}}\right)\right)}\,,\\
\frac{d \rho}{d \Psi}_\mathrm{B} (r) = &-\Bigg[r_{\text{NFW}}^3 \rho _s r^{1-\gamma _{\text{NFW}}} \left(r_{\text{NFW}}+r\right){}^{\gamma _{\text{NFW}}-4} \left(\gamma _{\text{NFW}} r_{\text{NFW}}+3
   r\right)\Bigg] \\&\Bigg/ \Bigg[G \Bigg(-M_{\text{BH}}+4 \pi  (-1)^{\gamma _{\text{NFW}}} r_{\text{NFW}}^3 \rho _s \left(B_{-\frac{r}{r_{\text{NFW}}}}\left(3-\gamma _{\text{NFW}},\gamma
   _{\text{NFW}}-2\right)-B_{-\frac{r_{\text{sp}}}{r_{\text{NFW}}}}\left(3-\gamma _{\text{NFW}},\gamma _{\text{NFW}}-2\right)\right) \\&+\frac{4 \pi  \rho _{\text{sp}}
   \left(r_{\text{sp}}^3 r_s^{\gamma _{\text{sp}}}-r_s^3 2^{3-\gamma _{\text{sp}}} r_{\text{sp}}^{\gamma _{\text{sp}}}\right) r_s^{-\gamma _{\text{sp}}}}{\gamma
   _{\text{sp}}-3}\Bigg)\Bigg]\,,
\end{split}
\end{equation}
and,
\begin{equation}
\begin{split}
\frac{d^2 \rho}{d \Psi^2}_\mathrm{A} (r) = &\Bigg[4^{\gamma _{\text{sp}}} \left(\gamma _{\text{sp}}-3\right){}^2 \gamma _{\text{sp}} \rho _{\text{sp}} r^{\gamma _{\text{sp}}+2} r_{\text{sp}}^{\gamma _{\text{sp}}}
   r_s^{2 \gamma _{\text{sp}}} \Bigg(M_{\text{BH}} 2^{\gamma _{\text{sp}}} \left(\gamma _{\text{sp}}-3\right) \left(\gamma _{\text{sp}}-1\right) \left(r
   r_s\right){}^{\gamma _{\text{sp}}} \\&+8 \pi  \rho _{\text{sp}} \left(4 r_s^3 \left(\gamma _{\text{sp}}-1\right) \left(r r_{\text{sp}}\right){}^{\gamma _{\text{sp}}}-r^3
   2^{\gamma _{\text{sp}}} \left(r_s r_{\text{sp}}\right){}^{\gamma _{\text{sp}}}\right)\Bigg) \Bigg] \Bigg/ \Bigg[G^2 \Bigg(M_{\text{BH}} 2^{\gamma _{\text{sp}}} \left(\gamma
   _{\text{sp}}-3\right)  \\&\times\left(r r_s\right){}^{\gamma _{\text{sp}}}+4 \pi  \rho _{\text{sp}} \left(8 r_s^3 \left(r r_{\text{sp}}\right){}^{\gamma _{\text{sp}}}-r^3
   2^{\gamma _{\text{sp}}} \left(r_s r_{\text{sp}}\right){}^{\gamma _{\text{sp}}}\right)\Bigg){}^3\Bigg]\,,\\
\frac{d^2 \rho}{d \Psi^2}_\mathrm{B} (r) = &\Bigg[r_{\text{NFW}}^3 \rho _s r^{2-\gamma _{\text{NFW}}} \left(r_{\text{NFW}}+r\right){}^{\gamma _{\text{NFW}}-5} \Bigg(G \left(8 r \gamma _{\text{NFW}}
   r_{\text{NFW}}+\gamma _{\text{NFW}} \left(\gamma _{\text{NFW}}+1\right) r_{\text{NFW}}^2+12 r^2\right)  \\&\times\bigg(-M_{\text{BH}}+4 \pi  (-1)^{\gamma _{\text{NFW}}}
   r_{\text{NFW}}^3 \rho _s \left(B_{-\frac{r}{r_{\text{NFW}}}}\left(3-\gamma _{\text{NFW}},\gamma
   _{\text{NFW}}-2\right)-B_{-\frac{r_{\text{sp}}}{r_{\text{NFW}}}}\left(3-\gamma _{\text{NFW}},\gamma _{\text{NFW}}-2\right)\right) \\&+\frac{4 \pi  \rho _{\text{sp}}
   \left(r_{\text{sp}}^3 r_s^{\gamma _{\text{sp}}}-r_s^3 2^{3-\gamma _{\text{sp}}} r_{\text{sp}}^{\gamma _{\text{sp}}}\right) r_s^{-\gamma _{\text{sp}}}}{\gamma
   _{\text{sp}}-3}\bigg)+r^3 \left(r_{\text{NFW}}+r\right) \left(\gamma _{\text{NFW}} r_{\text{NFW}}+3 r\right) \\&\times\Bigg(\bigg(2 G \Bigg(M_{\text{BH}}-4 \pi  (-1)^{\gamma
   _{\text{NFW}}} r_{\text{NFW}}^3 \rho _s \left(B_{-\frac{r}{r_{\text{NFW}}}}\left(3-\gamma _{\text{NFW}},\gamma
   _{\text{NFW}}-2\right)-B_{-\frac{r_{\text{sp}}}{r_{\text{NFW}}}}\left(3-\gamma _{\text{NFW}},\gamma _{\text{NFW}}-2\right)\right) \\&+\frac{\pi  \left(2^{5-\gamma
   _{\text{sp}}} \rho _{\text{sp}} r_{\text{sp}}^{\gamma _{\text{sp}}} r_s^{3-\gamma _{\text{sp}}}-4 r_{\text{sp}}^3 \rho _{\text{sp}}\right)}{\gamma
   _{\text{sp}}-3}\Bigg)\bigg) \bigg/ r^3 - 4 \pi  G r_{\text{NFW}}^3 \rho _s r^{-\gamma _{\text{NFW}}} \left(r_{\text{NFW}}+r\right){}^{\gamma
   _{\text{NFW}}-3}\bigg)\Bigg)\Bigg] \\& \Bigg/ \Bigg[G^3 \Bigg(-M_{\text{BH}}+4 \pi  (-1)^{\gamma _{\text{NFW}}} r_{\text{NFW}}^3 \rho _s
   \left(B_{-\frac{r}{r_{\text{NFW}}}}\left(3-\gamma _{\text{NFW}},\gamma _{\text{NFW}}-2\right)-B_{-\frac{r_{\text{sp}}}{r_{\text{NFW}}}}\left(3-\gamma
   _{\text{NFW}},\gamma _{\text{NFW}}-2\right)\right) \\&+\frac{4 \pi  \rho _{\text{sp}} \left(r_{\text{sp}}^3 r_s^{\gamma _{\text{sp}}}-r_s^3 2^{3-\gamma _{\text{sp}}}
   r_{\text{sp}}^{\gamma _{\text{sp}}}\right) r_s^{-\gamma _{\text{sp}}}}{\gamma _{\text{sp}}-3}\Bigg){}^3\Bigg]\,.
\end{split}
\end{equation}

\subsection{gNFW}

If we consider the gNFW density profile for DM, namely \Cref{eq:rho_gNFW}, and we disregard the effect from baryonic matter and Sgr $A^*$, then we can get the following expression for the mass from the center of the galaxy to a radial position, $r$:
\begin{equation}
\begin{split}
M (r) = -4 \pi  (-1)^{\gamma } \, \rho_s \, r_s^3 \,  B_{-r / r_s } \left( 3-\gamma ,\gamma -2 \right)\,.
\end{split}
\end{equation}
$\Psi$ takes the form:
\begin{equation}
\begin{split}
\Psi (r) = -&4 \pi \, G (-1)^{\gamma} \, \rho _s \, r_s^3 \left[\frac{\left(-\frac{R_{\max }}{R_{\max }+r_s}\right){}^{2-\gamma } -\left(-\frac{r}{r_s+r}\right){}^{2-\gamma }}{(\gamma -2) r_s} -\frac{B_{-R_{\max }/r_s}(3-\gamma ,\gamma -2)}{R_{\max }} +\frac{B_{-r/r_s}(3-\gamma ,\gamma -2)}{r}\right]\,.
\end{split}
\end{equation}

The derivatives are:
\begin{equation}
\begin{split}
\frac{d \rho}{d \Psi} (r) = &-\frac{r (-r)^{-\gamma } \left(r_s+r\right){}^{\gamma -4} \left(\gamma \,  r_s+3 \, r\right)}{4  \, \pi  \, G B_{-r / r_s} (3-\gamma ,\gamma -2)}\,,\\
\frac{d^2 \rho}{d \Psi^2} (r) = &\frac{(-1)^{-2 \gamma } r^{2-\gamma } \left(r_s+r\right){}^{\gamma -7} }{16 \pi ^2 G^2 r_s^3 \rho _s B_{-r / r_s}(3-\gamma ,\gamma -2){}^3}\,\\
&\times \left[\left(r_s+r\right){}^2 \left(6 r^2+6 (\gamma -1) r r_s+(\gamma -1) \gamma  r_s^2\right) B_{-r / r_s}(3-\gamma ,\gamma -2)+(-r)^{3-\gamma } \left(r_s+r\right){}^{\gamma } \left(\gamma  r_s+3 r\right)\right]\,.
\end{split}
\end{equation}

\bibliographystyle{apsrev4-1}
\bibliography{biblio}{}

\end{document}

%% file: custom_commands.tex
\newcounter{CommentCount}
\definecolor{MH}{rgb}{0.0,0.6,9}
\definecolor{palatinate}{rgb}{0.494, 0.192, 0.482}

\definecolor{teal}{HTML}{008080}

%% file: units.tex
\DeclareSIUnit \s {\second}
\DeclareSIUnit \ns {\nano\second}
\DeclareSIUnit \mus {\micro\second}
\DeclareSIUnit \ms {\milli\second}
\DeclareSIUnit \MB {\mega\byte}
\DeclareSIUnit \GB {\giga\byte}
\DeclareSIUnit \TB {\tera\byte}
\DeclareSIUnit \PB {\peta\byte}
\DeclareSIUnit \Mbps {\mega\bit/\s}
\DeclareSIUnit \Gbps {\giga\bit/\s}
\DeclareSIUnit \Tbps {\tera\bit/\s}
\DeclareSIUnit \Pbps {\peta\bit/\s}
\DeclareSIUnit \kton {\kilo\tonne} % changed  back to kton
\DeclareSIUnit \kt {\kilo\tonne}
\DeclareSIUnit \Mt {\mega\tonne}
\DeclareSIUnit \eV {\electronvolt}
\DeclareSIUnit \keV {\kilo\electronvolt}
\DeclareSIUnit \MeV {\mega\electronvolt}
\DeclareSIUnit \GeV {\giga\electronvolt}
\DeclareSIUnit \TeV {\tera\electronvolt}
\DeclareSIUnit \PeV {\peta\electronvolt}
\DeclareSIUnit \EeV {\exa\electronvolt}
\DeclareSIUnit \m {\meter}
\DeclareSIUnit \cm {\centi\meter}
\DeclareSIUnit \in {\inchcommand}
\DeclareSIUnit \km {\kilo\meter}
\DeclareSIUnit \kV {\kilo\volt}
\DeclareSIUnit \kW {\kilo\watt}
\DeclareSIUnit \MW {\mega\watt}
\DeclareSIUnit \MHz {\mega\hertz}
\DeclareSIUnit \mrad {\milli\radian}
\DeclareSIUnit \year {years}
\DeclareSIUnit \POT {POT}
\DeclareSIUnit \sig {$\sigma$}
\DeclareSIUnit\parsec{pc}
\DeclareSIUnit\lightyear{ly}
\DeclareSIUnit\foot{ft}
\DeclareSIUnit\ft{ft}
\DeclareSIUnit \ppb{ppb}
\DeclareSIUnit \ppt{ppt}
\DeclareSIUnit \samples{S}
\DeclareSIUnit \pe{PE}
\DeclareSIUnit \T{T}

\newcommand{\enu}{\E_\enu}